\documentclass[aps,prl,letterpaper,amsmath,amssymb,superscriptaddress,nofootinbib,twocolumn]{revtex4-1}
\usepackage{graphicx}
\usepackage[usenames]{color}
\usepackage[colorlinks=true,linkcolor=blue,citecolor=blue,anchorcolor=green,urlcolor=blue]{hyperref}
\usepackage{subfigure}
\usepackage{color}
\usepackage{mathrsfs}
\def\be{\begin{equation}}
\def\ee{\end{equation}}
\def\ba{\begin{aligned}}
\def\ea{\end{aligned}}
\usepackage[normalem]{ulem}

\usepackage{booktabs}
\usepackage[percent]{overpic}
\usepackage{tabularx}
\usepackage{multirow}
\usepackage{CJK}
\usepackage{orcidlink}
\usepackage{booktabs}
\usepackage{makecell}

\newcommand {\opc}{\hat{c}}
\newcommand {\ops}{\hat{S}}
\newcommand {\opn}{\hat{n}}

\begin{document}
\title{Intertwined spin and charge dynamics in one-dimensional supersymmetric t-J model}

\newtheorem{theorem}{Theorem}[section]
\newtheorem{lemma}[theorem]{Lemma}
\author{Yunjing Gao\orcidlink{0000-0002-1727-2577}}
\affiliation{Tsung-Dao Lee Institute,
Shanghai Jiao Tong University, Shanghai, 201210, China}

\author{Jianda Wu\orcidlink{0000-0002-3571-3348}}
\altaffiliation{wujd@tongji.edu.cn}
\affiliation{School of Physics Science and Engineering, Tongji University, Shanghai 200092, China}
\affiliation{Shanghai Branch, Hefei National Laboratory, Shanghai 201315, China}
\affiliation{Tsung-Dao Lee Institute,
Shanghai Jiao Tong University, Shanghai, 201210, China}
\date{\today}

\begin{abstract}
Following the Bethe ansatz we determine the dynamical spectra of the one-dimensional supersymmetric $t$-$J$ model.
A series of fractionalized excitations are 
identified through two sets of Bethe numbers.
Typical patterns in each set are found to yield wavefunctions containing elementary spin and charge carriers,
manifested as distinct boundaries of the collective excitations in the spectra of single electron Green functions.
In spin channels,
gapless excitations fractionalized into two spin and a pair of positive and negative charge carriers,
extending to finite energy as multiple continua.
These patterns connect to the half-filling limit where only fractionalized spinons survive.
In particle density channel,
apart from spin-charge fractionalization,
excitations involving only charge fluctuations are observed.
Furthermore, nontrivial Bethe strings encoding bound state structure appear in channels of reducing or conserving magnetization,
where spin and charge constituents can also be identified.
These string states contribute significantly even to the low-energy sector in the limit of vanishing magnetization.

\end{abstract}
\maketitle

\paragraph*{Introduction.---}
A central challenge in condensed matter physics is understanding how charge and spin degrees of freedom collectively influence strongly correlated systems \cite{fazekas,Auerbach,KAChao,doi:10.1126/science.235.4793.1196,PhysRevB.37.3759}.
Due to better controllability compared to higher-dimensional systems, 
one-dimensional (1D) models can offer reliable
analytical insights into this interplay. 
A prominent example is the Luttinger liquid theory, which has proven remarkably successful in describing a wide range of 1D systems \cite{tomonaga_remarks_1950,PhysRev.119.1153,PhysRevLett.47.1840,Haldane_1981,Giamarchi:2003ooa}. 
However, 
validity of the theory is restricted to the zero-energy limit as it is built on linearized dispersions.
In contrast,
the Bethe ansatz (BA) approach is advantageous in revealing energy- and momentum-resolved information across the full Brillouin zone.
Particularly,
it has been employed to investigate dynamical structure factors (DSFs) of magnetic systems \cite{PhysRevB.62.14871,Biegel_2002,doi:10.1143/JPSJ.73.3008,PhysRevLett.95.077201,Caux_2005,caux_correlation_2009,Kohno,PhysRevB.100.184406},
yielding results in excellent agreement with experimental data from real materials \cite{bera_dispersions_2020,PhysRevLett.123.067202}. 
However,
the computational complexity increases significantly when both spin and charge come into play.
The BA-solvable 1D supersymmetric (SUSY) $t$-$J$ model serves as an ideal playground here \cite{PhysRevB.12.3795,PhysRevB.36.5177,Sarkar_1990,PhysRevLett.64.2567,PhysRevB.46.9147,PhysRevB.44.130,PhysRevB.46.14624,FOERSTER1993611},
where progress was made a decade ago in obtaining its form factors \cite{Hutsalyuk_2016,Hutsalyuk_2017,HUTSALYUK2016902,Fuksa_2017},
thereby 
enabling the study of spin and charge dynamics in a rigorous fashion.

In this article,
after briefly reviewing BA solutions of the SUSY $t$-$J$ model,
we present detailed
spin and charge DSFs.
Featured Bethe number (BN) patterns are regarded as elementary constituents (referred to as particles).
Different combinations of them give rise to various fractionalized excitations in different operator channels.
Among these,
non-trivial string states are 
found to dominate the spectra beyond low-energy sector, 
which gradually share dominance with the low-lying real solutions when magnetization approaches zero.
Last, we discuss the
contribution of multi-particle states and evolution of the spectra with a series of electron fillings and magnetizations, where rich multi-spin and charge fractionalizations are revealed.

\paragraph*{The model and the Bethe wavefunctions.---}
The 1D supersymmetric $t$-$J$ model with periodic boundary condition in magnetic field is given by 
\be
\ba
\mathcal{H}&=-t \,\mathcal{P}\sum_{j=1,\sigma=\uparrow,\downarrow}^L\left(\opc_{j,\sigma}^\dagger \opc_{j+1,\sigma}+\opc_{j+1,\sigma}^\dagger \opc_{j,\sigma}\right)\mathcal{P}
\\
&+\sum_{j=1}^L\left(J\mathbf{\ops}_{j}\cdot\mathbf{\ops}_{j+1}-\frac{J}{4}\hat{n}_j\hat{n}_{j+1}-g\ops_j^z\right) \label{eq:tjham}
\ea
\ee
with $J=2t$ (set $t=1$ in the following), magnetic field $g$,  
electron creation and annihilation operators on $j$-th site $\opc_{j,\sigma}^\dagger$ and $\opc_{j,\sigma}$,
electron density $\opn_j$,
the projection operator $\mathcal{P}=\prod_{j=1}^L(1-\opn_{j,\uparrow}\opn_{j,\downarrow})$, and the spin operators $
\ops_i^+=\opc_{i,\uparrow}^\dagger \opc_{i,\downarrow},\,
\ops_i^-=\opc_{i,\downarrow}^\dagger \opc_{i,\uparrow},\,
\ops_i^z = \frac{1}{2}(\opn_{i,\uparrow}-\opn_{i,\downarrow}).
$
The eigenfunctions of $\mathcal{H}$ can be solved from the
nested BA equations \cite{FOERSTER1993611}
which involve two sets of rapidities $\{v_j\}$ and $\{\gamma_\alpha\}$,
with $j=1,\cdots, N_{hd} \equiv N_h+N_\downarrow$ and $\alpha=1,\cdots,N_h$.
$N_h$ and $N_\downarrow$ ($N_\uparrow$) label the number of holes and down (up) spin electrons.
We adopt the string hypothesis for $v_j$ to solve the equations \cite{Bethe,10.1143/PTP.46.401,FOERSTER1993611},
i.e., $v_{a j}^n=v_{a}^n+i(n+1-2j)$, $v_a^n\in\mathbb{R}$.
Here $j=1,\cdots, n$, $a=N_1,\cdots,N_n$ and $n=1,2,\cdots$ with the constraint $N_{hd}=\sum_{n}nN_n$.
In the following
states with real rapidities are solely denoted as $\mathscr{L}_1$.
String states containing a set of $\{v_{aj}^2\}$ or $\{v_{aj}^3\}$ apart from real $v_{aj}^1$'s are denoted as $\mathscr{L}_2$ or $\mathscr{L}_3$, respectively.
The rapidities can be solved from the BA equations \cite{FOERSTER1993611},
\be\ba
2\pi I_a^n&=L\theta\left(\frac{v_a^n}{n}\right)-\sum_m\sum_{\substack{b=1\\(m,b)\neq(n,a)}}^{N_m}\Theta_{nm}(X)
+\sum_{\beta=1}^{N_h}\theta\left(Y\right),\\
2\pi J_\beta &=\sum_n\sum_{a=1}^{N_n}\theta\left(Y\right),
\label{eq:logBAE}
\ea\ee
with $\theta(x)=2\arctan x$, $X = v_a^n-v_b^m$, $Y=(v_a^n-\gamma_\beta)/n$,
$\Theta_{n \neq m}(x)=\theta\left(x/|n-m|\right)+2\theta\left(x/(|n-m|+2)\right)+\cdots+2\theta\left(x/(n+m-2)\right)+\theta\left(x/(n+m)\right)$ and
$\Theta_{nn}(x)=2\theta\left(x/2\right)+2\theta\left(x/4\right)+\cdots+2\theta\left(x/(2n-2)\right)+\theta\left(x/(2n)\right)$.
$\{I_a^n\}$ and $\{J_\beta\}$ are Bethe numbers (BNs) used to solve one set of rapidities to yield one Bethe state.
$I_a^n$ are integers (half-integers) if $L+N_h-N_{n}$ is odd (even) and $J_\beta$ are integers (half-integers) if $\sum_nN_n$ is even (odd).
The BNs are restricted by
$|I_a^n|\leq \frac{1}{2}\left(L+N_h-\sum_{m}t_{mn}N_m-1\right)$
and $|J_\beta|\leq \frac{1}{2}\sum_{n}N_n-1$, with $t_{mn}=2\min(m,n)-\delta_{nm}$.
Energy and momentum of a Bethe state are given by $E=L-\sum_{j=1}^{N_{hd}}4/(1+v_j^2)$ and
$P=\sum_{j=1}^{N_{hd}}k_{j}=\sum_{j=1}^{N_{hd}}2\arctan(v_j)+\pi\mod(L-\sum_n N_{n}-1,2)$.

\begin{figure}[t]
\centering
    \includegraphics[width=0.45\textwidth]{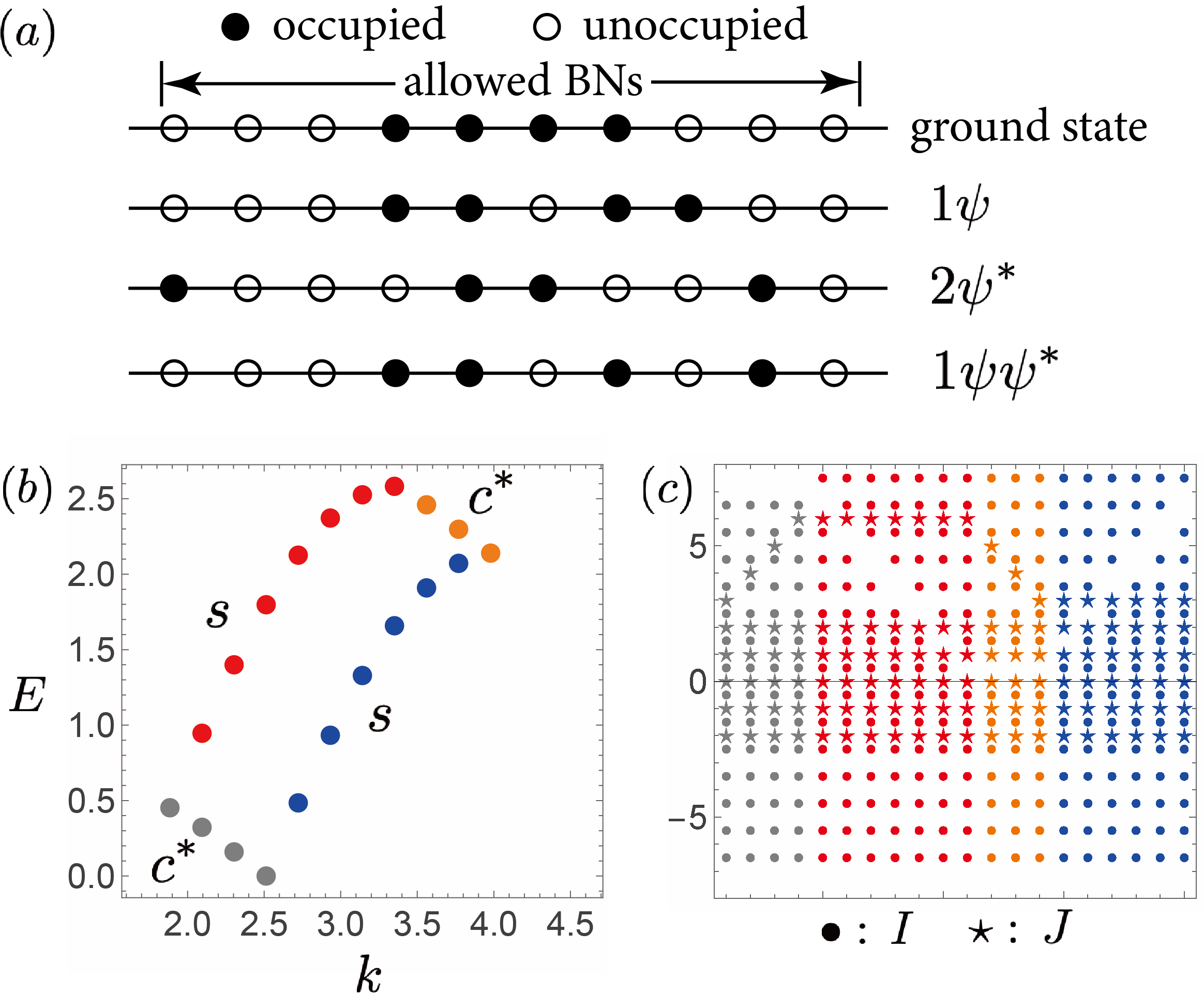}
\caption{($a$) Examples of BN patterns.
The ground state configurations for both $\{I_a^n\}$ and $\{J\}$
possess minimal total absolute value.
$n \psi$ are created by choosing desired number of BNs out of 
ground state occupations combined with $n$ innermost unoccupied BNs, either from the left, right or both sides.
$n \psi^*$ are created by replacing $n$ outermost occupied BNs by unoccupied BNs.
$n\psi\psi^*$ corresponds to moving $n$ occupied BNs to unoccupied positions.
These classifications can overlap with each other, e.g., $1\psi_c$ is a subset of $1\psi_c\psi_c^*$.
($b$,$c$) Examples of the energy and momentum of Bethe states ($b$) solved from BN patterns in ($c$). Each column in ($c$) stands for a set of BNs for one state (here $L=30$, $N_\downarrow=8$, $N_h=6$).
The colored dots and stars in ($c$) correspond to the states of the same color in ($b$), with the columns in ($c$) ordered according to the clockwise sequence of the points in ($b$) starting from the gapless point.
Part of the $c^*$ and $s$ bands are also illustrated.
\label{fig:BNpattern}}
\end{figure}

\begin{figure}[htp]
\centering
    \includegraphics[width=0.49\textwidth]{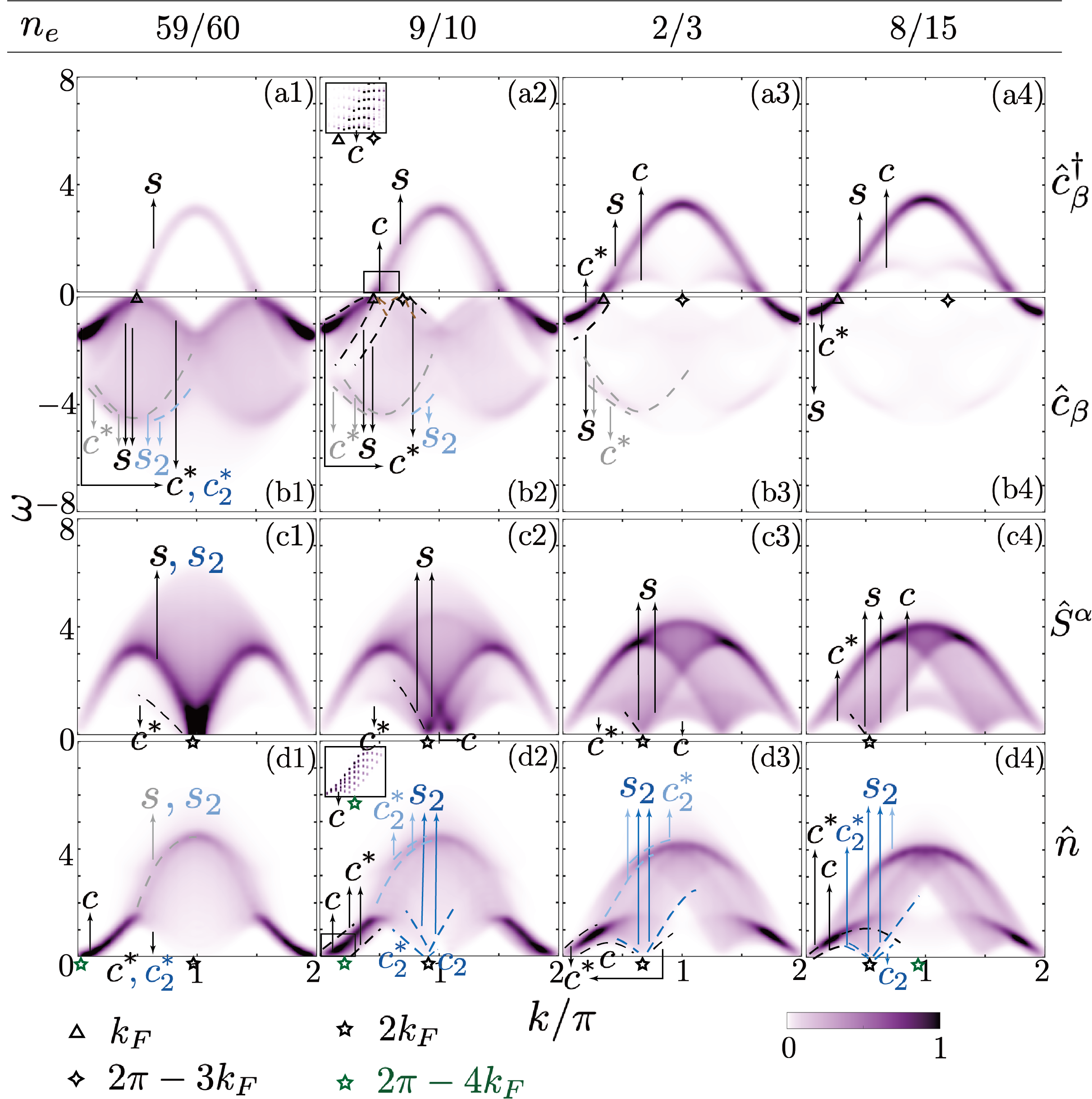}
    \caption{Dynamic structure factors at $g=0$ for different fillings in each column. 
    $\alpha=x,+,-$ and $\beta=\uparrow,\downarrow$ in the second through fourth columns.
    In the single hole case shown in the first column,
    $\alpha=z$ and $\beta=\uparrow$, with other channels exhibiting similarly as the SU$(2)$ symmetry is slightly broken.
    $k_F^\uparrow$ and $k_F^\downarrow$ differ by $\pi/L$ in this case, and we omit the difference in the denotations.
    Some featured single particle dispersions of $\mathscr{L}_1$ type are denoted by black and gray text and arrows, representing branches that reach or are gapped from the lowest energy in $\mathscr{L}_1$ region.
    The blue and light blue ones illustrate similarly for $\mathscr{L}_2$ case.    
    Markers on the horizontal axis illustrate the gapless points.}
    \label{fig:M0}
\end{figure}

\begin{figure}[htp]
\centering
    \includegraphics[width=0.46\textwidth]{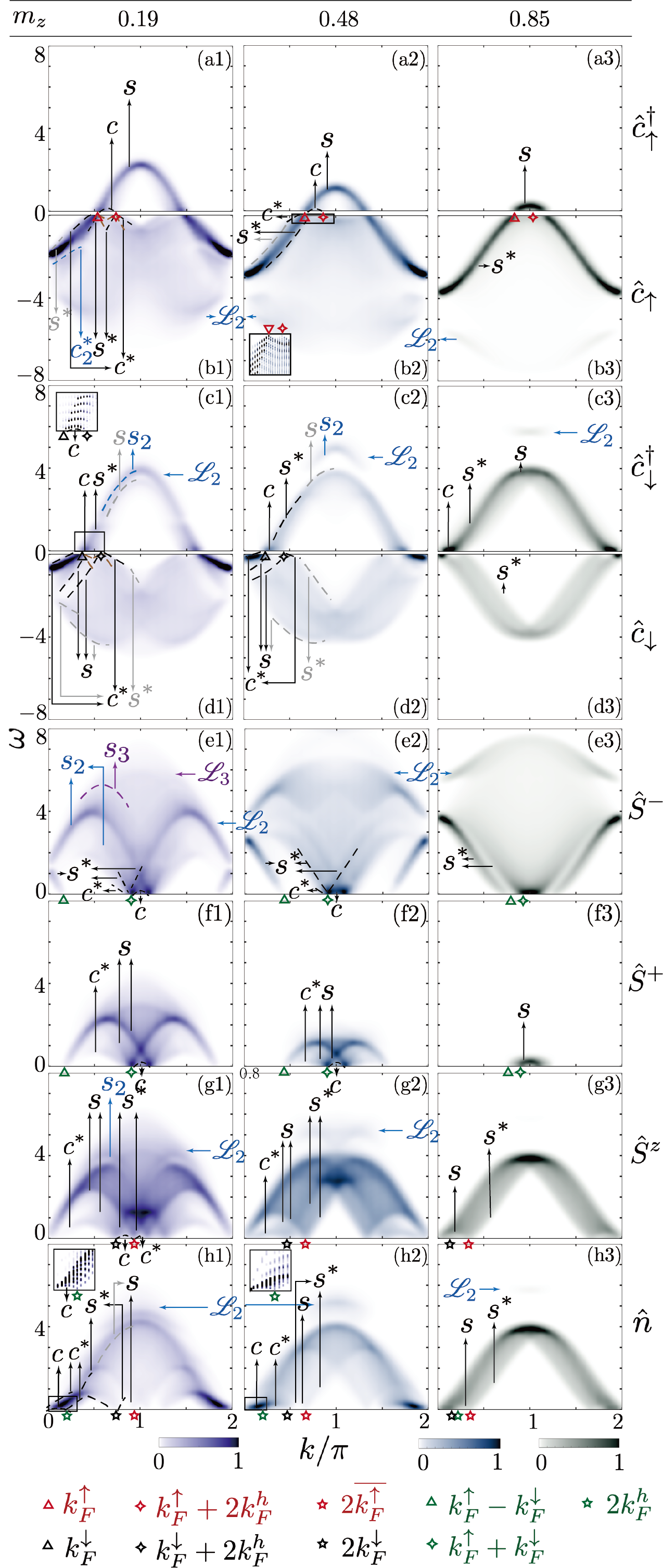}
    \caption{DSFs for $L=60$, $n_e=0.9$ ground states with different magnetization in each row.
    We define $\overline{k}\equiv\pi-k$.
    The same notations are applied as in Fig.~\ref{fig:M0},
    and purple color is used in $\mathscr{L}_3$ case.}
    \label{fig:N0p9}
\end{figure}

We focus on the DSF
$D({\hat{\mathcal{O}}};q,\omega)=2\pi\sum_{\nu}|\langle B(q+k_{\text{GS}},\omega_\nu)|\mathcal{O}_q|\text{GS}\rangle|^2\delta(\omega-\eta(\omega_\nu-\omega_{\text{GS}}-\mu\Delta N_e))$
where ground state $|\text{GS}\rangle$ carries momentum $k_{\text{GS}}$ and energy $\omega_{\text{GS}}$,
$\hat{\mathcal{O}_q} = (1/L)\sum_j \hat{\mathcal{O}_j}e^{-iq j} $, and
$|B(k,\omega_\nu)\rangle$ is a state with momentum $k$ and energy $\omega_\nu$. 
$\mu$ denotes the chemical potential and $\Delta N_e$ denotes electron number difference between the intermediate state and the ground state.
$\eta=-1$ for electron annihilation operator $\mathcal{O}$ and $\eta=1$ otherwise.
Bethe states with typical BN configurations are found dominant in the DSF.
We adopt the terminology psinon ($\psi$) and antipsinon ($\psi^*$) for spin models \cite{PhysRevB.66.054405,Kohno},
applying to both $\{I_a^n\}$ and $\{J_\beta\}$ and denoting as $\psi_s^n$ ($\psi_s^{*n}$) and $\psi_c$ ($\psi_c^*$), respectively [Fig.~\ref{fig:BNpattern}(a)].

\paragraph*{Multiple fractionalizations.---}
Spin and charge characteristics are manifested as different kinds of fractionalizations through different channels of the DSFs.
This section focuses on low energy sector and identifies these excitations through BN configurations.
We introduce $s_n$ band as the set of $\mathscr{L}_n$ states that involve moving $1\psi_s^1$ while keeping other BNs fixed,
The $s^*_n$, $c_n$ and $c^*_n$ bands are introduced analogously for the $1\psi^{*1}$, $1\psi_c$ and $1\psi_c^*$ cases, with the subscript $n=1$ omitted subsequently [Fig.~\ref{fig:BNpattern}(b,c)].
These basic patterns can be regarded as elementary constituents in the collective excitations.
In the thermodynamic limit,
$\{s, s^*\}$ and $\{c, c^*\}$ are characterized by two velocities (slope of the corresponding branches) near zero energy \cite{PhysRevB.46.14624, Giamarchi:2003ooa}.
For brevity, we refer to certain momenta and their $\pi$-symmetric counterparts interchangeably in the following discussion.

Adding an electron into the ground state leads to low energy excitations from Fermi point $k_F=n_e \pi/2$ to $k>k_F$ at $m_z=0$ ($n_e$ denotes electron density),
where fractionalized excitations $1\psi_s1\psi_c$ composed by $s$ and $c$ can be identified at finite doping [Fig.~\ref{fig:M0} (a1-a4)].
In the presence of magnetic field,
the up and down spin fermi points split into $k_F^\uparrow=n_e(1+m_z)\pi/2$ and $k_F^\downarrow=n_e(1-m_z)\pi/2$,
with magnetization $m_z=(N_\uparrow-N_\downarrow)/(L-N_h)$.
Adding an up spin electron leads to $1\psi_s1\psi_c$ continuum [Fig.~\ref{fig:N0p9} (a1-a3)],
while in $D(\opc_\downarrow^\dagger)$ the low energy part is dominated by $1\psi_s^*1\psi_c$ [Fig.~\ref{fig:N0p9} (c1-c3)].
In both cases,
the upper and lower boundaries follow different ``single (quasi)particle" dispersions which can be understood as moving either $s$ ($s^*$) or $c$ particles, 
while keeping the other at rest.
Removing an up spin electron is reflected as dominant $1\psi_s^*1\psi_c^*$ excitations within $k<k_F^\uparrow$ [Fig.~\ref{fig:N0p9} (b1-b3)],
parallel to $1\psi_s1\psi_c^*$ within $k<k_F^\downarrow$ in the down spin case [Fig.~\ref{fig:N0p9} (d1-d3)].
The two cases converge at $g=0$ (zero field) appearing as $1\psi_s1\psi_c^*$ continuum [Fig.~\ref{fig:M0} (b1-b4)].
Additional gapless points and finer structure can also be observed,
which encodes multiple process and is deferred to the section of \textit{Particle-hole pairs}.

Meanwhile,
spin-$1$ excitations are also fractionalized.
Gapless spin flip involves scattering an up-spin at $\pm k_F^\uparrow$ to a down-spin at $\pm k_F^\downarrow$ (or vice versa),
resulting in momentum transfers of $\pm k_F^\uparrow \pm k_F^\downarrow$.
For $D(\hat{S}^-)$ the fractional constituents are implied by the four explicit dispersions,
$c^*$, $c$ and two $s^*$'s starting from $k_F^\uparrow+k_F^\downarrow$ with $\omega=0$.
In addition to single particle branches,
dominant spectral weight comes from varying momenta of two particles out of the four,
namely, the $2\psi_s^*$, $1\psi_s^*1\psi_c$, $1\psi_s^*1\psi_c^*$ and $1\psi_c\psi_c^*$ continua [Fig.~\ref{fig:N0p9} (e1-e3)].
On the contrary, in $D(\ops^+)$
starting from $k_F^\uparrow+k_F^\downarrow$ at zero energy,
$c^*$, $c$  and two $s$ bands can be identified,
and the collective two-particle combinations follow [Fig.~\ref{fig:N0p9} (f1-f3)].

\paragraph*{Particle-hole pairs.---}
We begin with $D(\ops^z)$.
Perfect nesting scatterings appear in both spin up and down Fermi surfaces,
leading to gapless excitations at $2k_F^\uparrow$ and $2k_F^\downarrow$ [Fig.~\ref{fig:N0p9} (g1-g3)].
In connection with half-filling limit without external field,
the significant spectral weight near $k=\pi$ reflects strong antiferromagnetic correlations [Fig.~\ref{fig:M0} (c1)],
in contrast to vanishing spectra weight at $k=0$.
The $c$, $c^*$, $s$ and $s^*$ dispersions can be identified sprouting from $2k_F^{\uparrow/\downarrow}$,
accompanied by continuum region dominated by pairwise combinations.
And for zero field,
$2k_F^\uparrow$ and $2k_F^\downarrow$
together with corresponding continua merge together [Fig.~\ref{fig:M0} (c1-c4)].

Next, we discuss $D(\opn)$ which also preserves particle number and magnetization as the $\ops^z$ channel.
$D(\opn)$ involves gapless particle-hole excitations from up/down-spin and hole Fermi surfaces
with transfer momenta $k=0$, $2k_F^\uparrow$, $2k_F^\downarrow$ and $2k_F^h=2\pi-4k_F$ [Fig.~\ref{fig:N0p9} (h1-h3)].
Analogous to the $D(\hat{S}^z)$,
the excitations at $2k_F^{\uparrow/\downarrow}$ exhibit fractionalizations that involve both spin and charge degrees of freedom.
Besides,
significant $1\psi_c\psi_c^*$ continuum touches $k=0$ and $2k_F^h$ at zero energy, which is contributed by charge degree of freedom solely, 
as opposed to $D(\ops^z)$.

The identification of hole Fermi surface excitations provides a further understanding in other channels.
In $D(\opc_{\uparrow/\downarrow})$,
gapless points $k_F^{\uparrow/\downarrow}+2k_F^h$ ($3k_F$ in the zero field limit) exist apart from $k_F^{\uparrow/\downarrow}$[Fig.~\ref{fig:N0p9} (b, d)].
Excitations associated with the former and latter can be regarded as composite processes involving the removal of a fractionalized electron, together with a nesting $k=2k_F^h$ and $k=0$ scattering on the hole Fermi surface.
This provides a microscopic origin for the $3k_F$ anomaly in Ref.~\cite{PhysRevB.83.205113}.
Consequently,
besides $s$ (or $s^*$) and $c^*$ bands mentioned in the previous paragraph,
$c$ and another $c^*$ bands with opposite velocity can be found
near zero energy at both $k_F^{\uparrow/\downarrow}$ and $k_F^{\uparrow/\downarrow}+2k_F^h$ (brown dashed lines in Fig.~\ref{fig:M0} (b2) and Fig.~\ref{fig:N0p9} (b1, d1)).
It is worth noticing that in $D(\ops^z)$,
the two gapless points $2\overline{k_F^\uparrow}$ and $2k_F^\downarrow$ are separated by $2k_F^h$.
The difference in their spectral weight can be understood from the distinct behavior of the associated hole scattering [Fig.~\ref{fig:N0p9} (g1-g3)].

\paragraph*{String contribution.---}
Apart from the real Bethe solutions ($\mathscr{L}_1$) discussed previously, non-trivial string states ($\mathscr{L}_{n \ge 2}$) also contribute significantly to the $\opc_\downarrow^\dagger$, $\opc_\uparrow$, $\ops^-$, $\ops^z$, and $\opn$ channels. 
A common feature for finite magnetization is that as $m_z$ decreases, 
the string states reach lower energy with increasing spectral contribution [see the SM for individual contributions of $\mathscr{L}_n$ states].
Viewing from the spectral weight distribution,
distinct bands are preserved across different $\mathscr{L}_n$ regions,
and explicit boundaries of different $\mathscr{L}_n$ continua align,
particularly at low magnetization
[Fig.~\ref{fig:N0p9} (b, c, e, g, h)].

Explicitly,
for $D(\opc_\downarrow^\dagger)$
the dominant $s^*$ band extending from zero energy,
continues to higher energy with coexisting $s$ and $s_2$ [Fig.~\ref{fig:N0p9} (c1)].
Analogous to $1\psi_s1\psi_c$ fractionalization for $\mathscr{L}_1$ states,
$1\psi_s^11\psi_c$ dominates the $\mathscr{L}_2$ continuum.
For $D(\opc_\uparrow)$,
the $\mathscr{L}_2$ continuum has a lower boundary $c^*_2$ near $k=0$ aligning to an $\mathscr{L}_1$ edge,
as well as an explicit $1\psi_c^*1\psi_s^1$ region near the top [Fig.~\ref{fig:split} (a2, c2, h2), Fig.~\ref{fig:N0p9} (b1-b3)].
It's worth noting that in $D(\opc_\uparrow)$ and $D(\ops^-)$ at low $m_z$,
the explicit single-particle dispersions with different string lengths,
$s^*$ and $s_2$,
tend to connect continuously [Fig.~\ref{fig:N0p9} (b1, e1)].
For $D(\ops^-)$,
dominant contribution in the $\mathscr{L}_2$ continuum includes $1\psi_s^11\psi_s^2$, $1\psi_s^11\psi_c$, $1\psi_s^21\psi_c$ and $2\psi_s$.
Furthermore, $\mathscr{L}_3$ continuum can be observed above $\mathscr{L}_2$ with small magnetization.
In $D(\ops^z)$, 
an explicit $s_2$ lower boundary of the $\mathscr{L}_2$ continuum aligns to upper edge of $1\psi_s\psi_s^*$ [Fig.~\ref{fig:N0p9} (g1-g3)].
On the other hand,
$\mathscr{L}_2$ states contribute to $D(\opn)$
across the $\mathscr{L}_1$ dome ($s$) near $k=\pi$ with dominant $1\psi_s^11\psi_c^*$  [Fig.~\ref{fig:split} (b2, f2), Fig.~\ref{fig:N0p9} (h1-h3)].

\begin{figure}[t]
    \centering
    \includegraphics[width=0.46\textwidth]{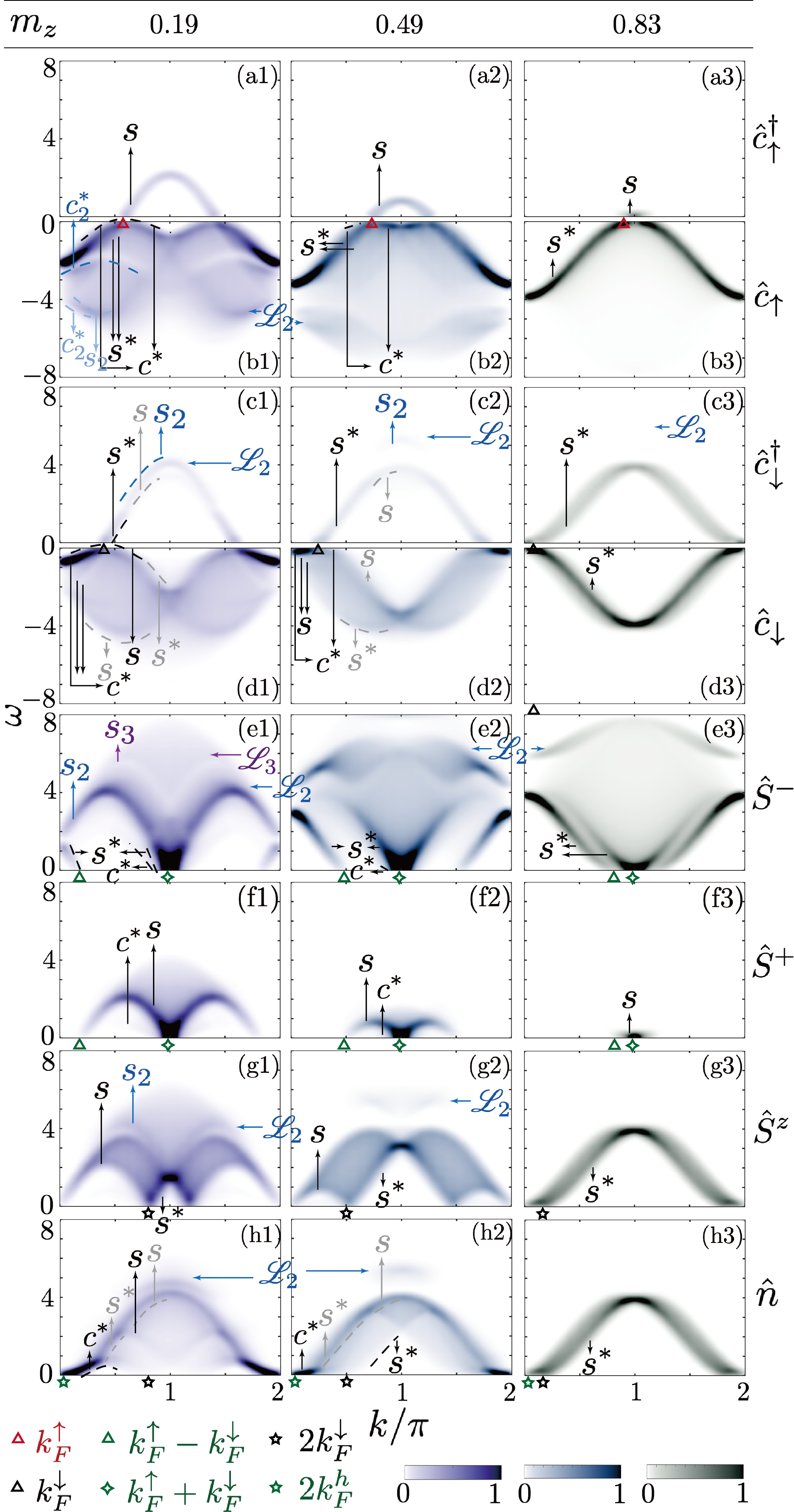}
    \caption{DSFs for ground state with one hole and $L=60$. Ground state with the same magnetization is illustrated by the same color. 
    The same notations are applied as in Figs.~(\ref{fig:M0},\ref{fig:N0p9}).
    Closed gapless points separated by $k=2\pi/L$ is illustrated as one point in the notations for brevity.
    \label{fig:1hole}}
\end{figure}

\paragraph*{Multi-particle states and spectra evolution.---}
Considering the full spectrum,
multi-particle states beyond two-particle ones also play an important role.
These states appear mainly in two ways.
The first case is indicated by multiple branches emerging from the gapless points,
which involve movement of more than two particles [e.g., Fig.~\ref{fig:N0p9} (e1)].
Second,
it appears as an extension of two-particle states to broader energy and momentum,
e.g., after $1\psi^*$ touches the boundary, another $1\psi$ is excited [Fig.~\ref{fig:N0p9} (c1)].
Among these, BN sets with substantial contributions are summarized in TABLE~\ref{tab:BN set}.

\begin{table}[h]
\centering 
\caption{Dominant BN sets in the DSFs.}
\begin{tabular}{ccccc}
    \hline 
    $\mathcal{O}$&$\opc_\uparrow^\dagger$&$\opc_\downarrow^\dagger$&$\opc_\uparrow$&$\opc_\downarrow$\\
    \hline
    BN&$1\psi_s1\psi_c$&\makecell{$1\psi_s\psi_s^*1\psi_c$\\$1\psi_s^11\psi_s^{2*}1\psi_c$}&\makecell{$1\psi_s\psi_s^*1\psi_c2\psi_c^*$\\$1\psi_s^11\psi_s^{2*}1\psi_c^*$}&\makecell{$1\psi_c\psi_c^*1\psi_s\psi_s^*$}\\
    \hline
    $\mathcal{O}$&$\ops^+$&$\ops^-$&$\ops^z$&$\opn$\\
    \hline BN&\makecell{$2\psi_s1\psi_c\psi_c^*$}&\makecell{$1\psi_s2\psi_s^*1\psi_c\psi_c^*$\\$2\psi_s^11\psi_s^{*2}1\psi_c^*$\\$2\psi_s^11\psi_s^{*3}1\psi_c^*$}&\makecell{$2\psi_s\psi_s^*1\psi_c\psi_c^*$\\$2\psi_s^11\psi_s^{*2}1\psi_c^*$}&\makecell{$2\psi_s\psi_s^*1\psi_c\psi_c^*$\\$2\psi_s^11\psi_s^{*2}1\psi_c^*$}\\
    \hline
\end{tabular}
\label{tab:BN set}
\end{table}

Next we discuss the evolution of DSF spectrum {\it{vs.}} particle density and magnetization. The momentum ranges for complete $s$ and $s^*$ bands follow $W_s=2\pi-\pi n_e(1+m_z)$ and $W_{s^*}=2\pi n_e m_z$.
And for $c$ and $c^*$ ones,
$W_c=2\pi(1-n_e)$ and $W_{c^*}=n_e\pi(1-m_z)$.
As $n_e$ increases,
there is less space to create electrons, 
leading to shrinked $\opc^\dagger$ spectra,
consistent with smaller $W_c$ here.
Similarly,
larger $m_z$ leads to suppressed up-spin creation  i.e., $D(\ops^+)$,
also indicated by smaller $W_s$.

Featured spectral weight transfer can be observed.
Consider spectrum evolution with decreasing $n_e$.
$\opc^\dagger$ maintains an explicit band-like shape accompanied with a continuum [Fig.~\ref{fig:M0} (a1-a4)].
In contrast, for $\opc$,
the spectra weight of the continuum distributed across $k=\pi$ gradually transfers to the band-like region near $k=0$ [Fig.~\ref{fig:M0} (b1-b4)].
For spin channels,
the incommensurate gapless points move towards $k=0$,
leading to more separated collective excitations for $k>\pi$ and $k<\pi$,
which meet at $\pi$ and form a strong region at finite energy [Fig.~\ref{fig:M0} (c1-c4)].
Furthermore,
collective excitations involving $1\psi_c$ and/or $1\psi_c^*$ are enhanced.
In $\hat n_e$ channel,
the pure charge response region $1\psi_c\psi_c^*$ is broadened with
decreasing $n_e$ [Fig.~\ref{fig:M0} (d1-d4)].
The continuum involving both spin and charge near $k=\pi$ further splits, 
evolving towards band-like shape with increasing spectral weight near $k=\pi$ at the upper boundary.
Either with small $n_e$ or large $m_z$,
both spin and charge spectra tend to a (broadened) band-like shape.
Consider increasing $m_z$ [Fig.~\ref{fig:N0p9}, Fig.~\ref{fig:1hole}].
Near gapless points, 
explicit suppression of the velocities of the $c$, $c^*$ particles can be observed from the slope of the branches.
In $D(\opc_\uparrow)$,
spectral weight of the continuum transfers to the boundaries,
while in $\opc_\downarrow$ branch it narrows into a band-like shape. 
In $\ops^z$ and $\opn$ channels the spectra evolve towards similar continuum shape with $s$ and $s^*$ as boundaries.

\paragraph*{Discussions.---}
At finite magnetization, adding or removing an electron reveals distinct behaviors. 
From the gapless points,
the $s$ and $s^*$ particles can be understood as carrying $S^z=1/2$ and $-1/2$, respectively,
while $c$ and $c^*$ can be attached charge $-1$ and $+1$, respectively.
Branches emerging from the same gapless points together imply the spin and charge properties in the continuum,
which is consistent with the observations in other channels.
For zero magnetic field,
given the limited space of unoccupied BNs for $\{I_a^n\}$,
we do not distinguish $s$ and $s^*$ but only regard it as a spin-$1/2$ object.

A gap $\Delta=g$ opens in $\ops^-$ at $k=0$, corresponding to Larmor precession modes \cite{PhysRevB.100.184406}.
Finite energy separation from $\omega=0$ can also be observed in the $\opc_\uparrow$ and $\opc_\downarrow$ channels at $k=0$ with $\Delta_{\uparrow} =(\mu-g)/2$ and $\Delta_{\downarrow} =(\mu+g)/2$,
with the difference reflecting the Zeenman splitting.
On the other hand,
in finite hole doping cases,
a region near $k=\pi$ with finite energy above the lower boundary contributes strong spectra weight in the spin channels [Fig.~\ref{fig:M0} (c2-c4), Fig.~\ref{fig:N0p9} (e, f, g)].
This arises from a high density of states where two continua cross, originating from gapless points at $k<\pi$ and $k>\pi$ with left- and right-moving $s$ ($s^*$) particles.

The 
configuration $N_\uparrow=N_\downarrow=N_h=1/3$  leads to a merge of Fermi points. 
For $\opn$ channel, the overlap of $2\pi-4k_F$ with $2k_F$ brings together the charge-only and spin-charge continua, yielding similar shapes in $\ops^z$ and $\opn$ [Fig.~\ref{fig:M0} (c3, d3)]. This similarity holds for $n_e<2/3$, reflecting an enhanced spectral weight 
involving both of spin and charge fluctuations in $D(\opn)$.

\paragraph*{Acknowledgments---}
We thank R. Yu, Y. Jiang, W. Ku, Z. Han and J. Yang for helpful discussions.
The work is sponsored by the National Natural Science Foundation of China Nos. 12450004, 12274288,  
the Innovation Program for Quantum Science and Technology Grant No. 2021ZD0301900. J.W. acknowledges the hospitality of Wilczek Quantum Center at Shanghai Institute for Advanced Studies of University of Science and Technology of China.

\bibliography{tjBA}

\clearpage
\appendix
\newpage
\onecolumngrid
\setcounter{figure}{0}
\makeatletter
\renewcommand{\thefigure}{S\@arabic\c@figure}
\setcounter{equation}{0} \makeatletter
\renewcommand \theequation{S\@arabic\c@equation}
\section*{supplementary material}

\section{Spectra from individual $\mathscr{L}_n$ states}
\begin{figure}[h!]
\centering
    \includegraphics[width=0.9\textwidth]{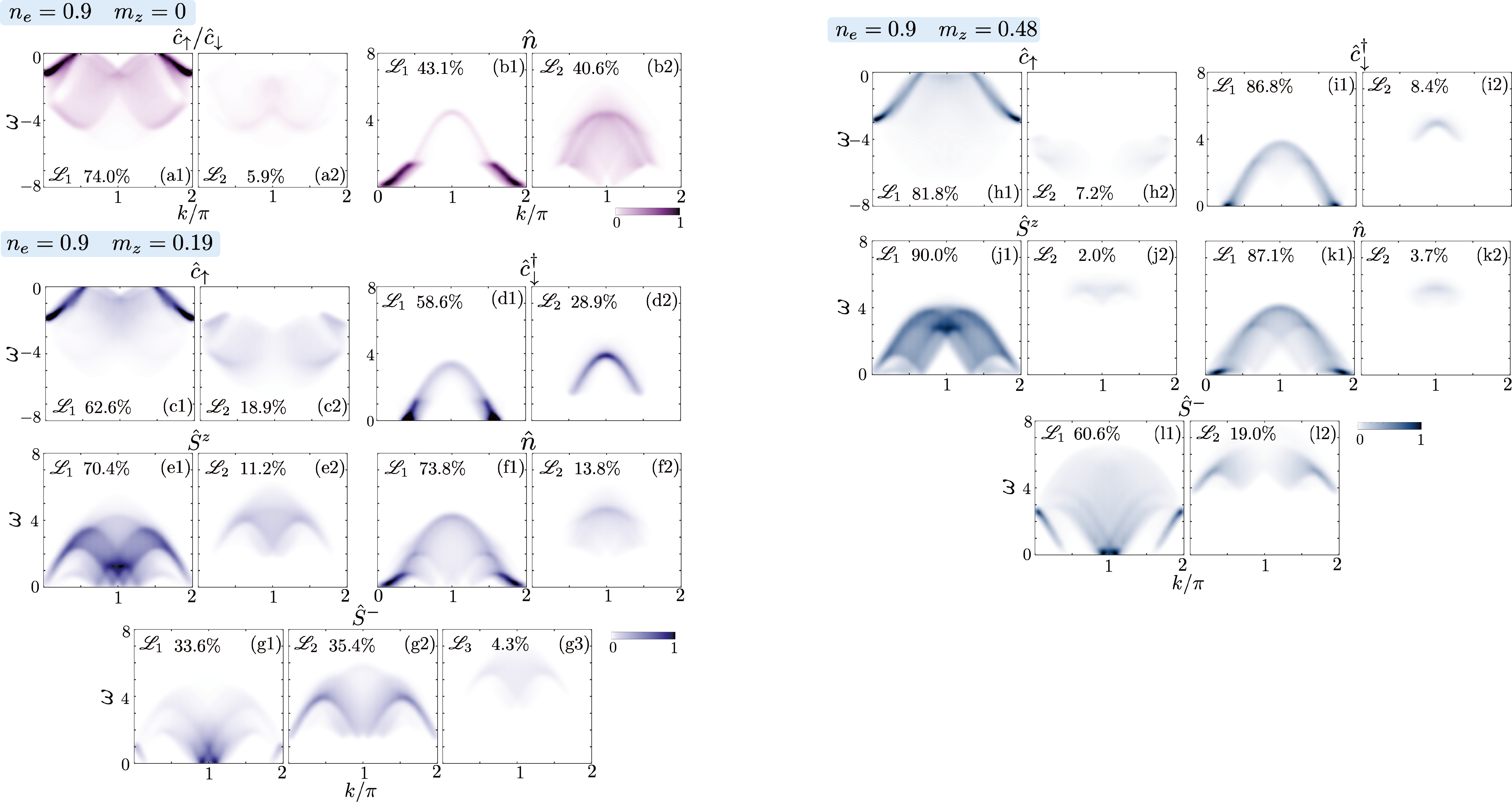}
    \caption{Example of DSFs contributed from $\mathscr{L}_n$ states with size $L=60$.
    Percentage in each figure shows the zero sum saturation of spectral weight contributed by $\mathscr{L}_n$ in the total spectrum.}
    \label{fig:split}
\end{figure}

As magnetization approaches zero,
the $\mathscr{L}_{n>1}$ states contribute in the low-energy sector,
and strongly overlap with $\mathscr{L}_1$ region.
To further clarify their contribution,
we split the spectra into different $\mathscr{L}_n$ parts shown in Fig.~\ref{fig:split}.
Contribution of $\mathscr{L}_n$ states are counted as the ratio of summation of corresponding spectral weight,
to the zero sum $T(\mathcal{O})=\sum_{q,\omega}D(\mathcal{O};q,\omega)$ in each channel. 
After excluding the vacuum contribution, for electron creation and annihilation channels
$T(\opc_{\uparrow/\downarrow}^\dagger)=N_h$ and
$T(\opc_{\uparrow/\downarrow})=N_{\uparrow/\downarrow}$.
For spin and charge densities,
$T(\ops^z)=(N_\uparrow+N_\downarrow)/4-(N_\uparrow-N_\downarrow)^2/(4L)$,
and $T(\opn)=N_h-N_h^2/L$.
At last,
$T(\ops^+)=N_h/2+N_\downarrow$ and $T(\ops^-)=N_h/2+N_\uparrow$.
In summary,
for finite $m_z$,
$\mathscr{L}_{n\geq2}$ contribution increases with decreasing magnetization.

\end{document}